\documentclass[preprint,12pt]{elsarticle}

\usepackage{graphicx}
\usepackage{color}

\usepackage{amssymb,amsbsy,amsmath}

\newcommand {\mofe} {Mo$_{72}$Fe$_{30}$}

\def\bra#1{\langle \, {#1} \, | \,}
\def\ket#1{\, | \, {#1} \, \rangle}
\newcommand{\op}[1]{%
    \fontdimen12\textfont3=2pt\fontdimen12\scriptfont3=1.4pt%
    \!\null\mathop{\vphantom{#1}\smash{#1}}\limits_{\sim}\null\!}
\newcommand{\vecop}[1]{%
    \fontdimen12\textfont3=2pt\fontdimen12\scriptfont3=1.4pt%
    \!\null\mathop{\textbf{\vphantom{#1}\smash{#1}}}\limits_{\sim}\null\!}

\newcommand{\xref}[1]{\protect\ref{#1}}
\newcommand{\figref}[1]{Fig.~\protect\ref{#1}}

\def\bra#1{\langle \, {#1} \, | \,}
\def\ket#1{\, | \, {#1} \, \rangle}
\newcommand{\braket}[2]{\langle \, {#1} \, | \, {#2} \, \rangle}

\journal{Polyhedron}

\begin{document}

\begin{frontmatter}



\title{Advanced quantum methods for the largest magnetic molecules}


\author{J{\"u}rgen Schnack\corref{cor1}\fnref{bi}}
\ead{jschnack@uni-bielefeld.de}
\cortext[cor1]{corresponding author}
\author{J{\"o}rg Ummethum\fnref{bi}}
\address[bi]{Dept. of Physics, Bielefeld University, P.O. box
  100131, D-33501 Bielefeld, Germany}

\begin{abstract}
We discuss modern numerical methods for quantum spin systems
and their application to magnetic molecules.
\end{abstract}


\end{frontmatter}


\section{Introduction}
\label{sec-1}

The knowledge of the energy eigenvalues and eigenstates of small
magnetic systems such as magnetic molecules is indispensable for
a complete understanding of their spectroscopic, dynamic, and
thermodynamic properties. In this respect the numerical exact
diagonalization of the appropriate quantum Hamiltonian is the
method of choice. Nevertheless, such an attempt is very often
severely 
restricted due to the huge dimension of the underlying Hilbert
space. For a magnetic system of $N$ spins of spin quantum number
$s$ the dimension is $(2s+1)^N$ which grows exponentially with
$N$. 

Group theoretical methods can help to ease this numerical
problem. Along these lines much effort has been put into the development
of an efficient numerical diagonalization technique of the
Heisenberg model 
\begin{equation}
   \op{H}_\text{Heisenberg} = -2 \sum_{i<j} J_{ij}\, \vecop{s}(i)
   \cdot \vecop{s}(j)  
\label{eq-1-1}
\end{equation}
using irreducible tensor operators and thus
employing SU(2) symmetry of angular
momenta
\cite{GaP:GCI93,BCC:IC99,BeG:EPR,Tsu:group_theory,Tsu:ICA08,BBO:PRB07}. 
A combination of this meanwhile well established technique with
point-group symmetries could be developed over the past years,
first for those point-group symmetries that
are compatible with the spin coupling scheme, i.e. avoid
complicated basis transforms between different coupling
schemes \cite{DGP:IC93,Wal:PRB00,BOS:TMP06,SBO:JPA07}, later
also for general point groups \cite{ScS:PRB09,ScS:P09,ScS:IRPC10}.
Nevertheless, if the dimension of the largest Hilbert subspace
exceeds about $10^{5}$ complete numerically exact
diagonalization is no longer possible with current computers and
programs.  

Fortunately, a few very accurate approximations have been
developed that can be applied to quantum spin systems. In this
article we are going to discuss the Finite Temperature Lanczos Method
(FTLM), the Density Matrix Renormalization Group (DMRG) and its
dynamical variant as well as Quantum Monte Carlo (QMC).

For problems with
Hilbert space dimensions of up to roughly $10^{10}$ -- the
Finite Temperature Lanczos Method
(FTLM) provides a very accurate and astonishingly easy to program
method \cite{PhysRevB.49.5065,JaP:AP00}. In recent publications
we demonstrated that this method is indeed capable of evaluating
thermodynamic observables for magnetic molecules with an accuracy
that is nearly indistinguishable from exact
results \cite{ScW:EPJB10,GDM:DT11,HSP:ACIE12}. So far we
encountered only one problem where achieving a satisfying
accuracy posed a problem -- the Fe${}^{\text{III}}_{10}$ ferric wheel
\cite{Garlatti:PC}. Mathematically this methods relies on the
idea of trace estimators \cite{Hut:CSSC89}; it is not restricted
to spin systems and for instance also applied in quantum
chemistry \cite{MHL:CPL01,HLM:JCP02}.

The Density Matrix Renormalization Group
(DMRG) method is a variational method that approximates the true
eigenstates by so-called matrix-product states
\cite{Whi:PRL1992,Whi:PRB93,WhD:PRB93B,Sch:RMP05}. These states
are iteratively constructed, thus the method allows to treat the
full Heisenberg Hamiltonian but in a reduced Hilbert space. The Hilbert space is
truncated in a controlled way and the accuracy of the  method
can be estimated with the help of a truncation error. In the field of molecular magnetism DMRG
has been applied for instance to the Heisenberg icosidodecahedron
with $s=5/2$ \cite{ExS:PRB03,USL:JMMM13}, i.e. a model of the \mofe\
Keplerate \cite{MSS:ACIE99,MLS:CPC01}.

DMRG can be extended in order to evaluate transition matrix
elements. This variant is called Dynamical DMRG (DDMRG); it aims
at the calculation of dynamical correlation functions as needed
for the description of Inelastic Neutron Scattering cross
sections \cite{KuW:PRB99,Jec:PRB02}. In a recent article we
could show that DDMRG is able to model INS spectra of the very large
magnetic ring molecule Fe$_{18}$ with unprecedented accuracy and thus
allows to determine model parameters which would be impossible
using only observables such as susceptibility \cite{UNM:PRB12}.
This extension is not discussed in this article. 

Finally we would like to provide examples for the application of
Quantum Monte Carlo (QMC) \cite{SaK:PRB91,San:PRB99,San:AIPCP10}
to magnetic molecules. This approximate method which works
accurately only for non-frustrated quantum spin systems
\cite{HeS:PRB00}, has already been applied to several molecular
systems by Larry Engelhardt
\cite{EnL:PRB06,EMP:ACIE08,EMP:PRB09,TMB:CC09}.  He also provides a very
popular program -- FIT-MART -- with which one can deduce
Heisenberg exchange parameters from susceptibility data
\cite{EnR:FITMART10}. 

Some of the discussed methods are freely available as program
packages. Besides FIT-MART the program MAGPACK \cite{BCC:JCC99}
can be used for complete diagonalization. Approximate methods
such as DMRG and QMC are provided by the ALPS package
\cite{ALPS:JMMM07,PRV:PRE04,AWT:PRE05}. 

The article is organized as follows. In Section \xref{sec-2} the
Finite-Temperature Lanczos Method is introduced. Section
\xref{sec-3} discusses the application of Quantum Monte Carlo,
and Section \xref{sec-4} briefly introduces to the Density Matrix
Renormalization Group.

\section{Application of the Finite-Temperature Lanczos Method to
giant gadolinium clusters}
\label{sec-2}

\begin{figure}[ht!]
\centering
\includegraphics*[clip,width=55mm]{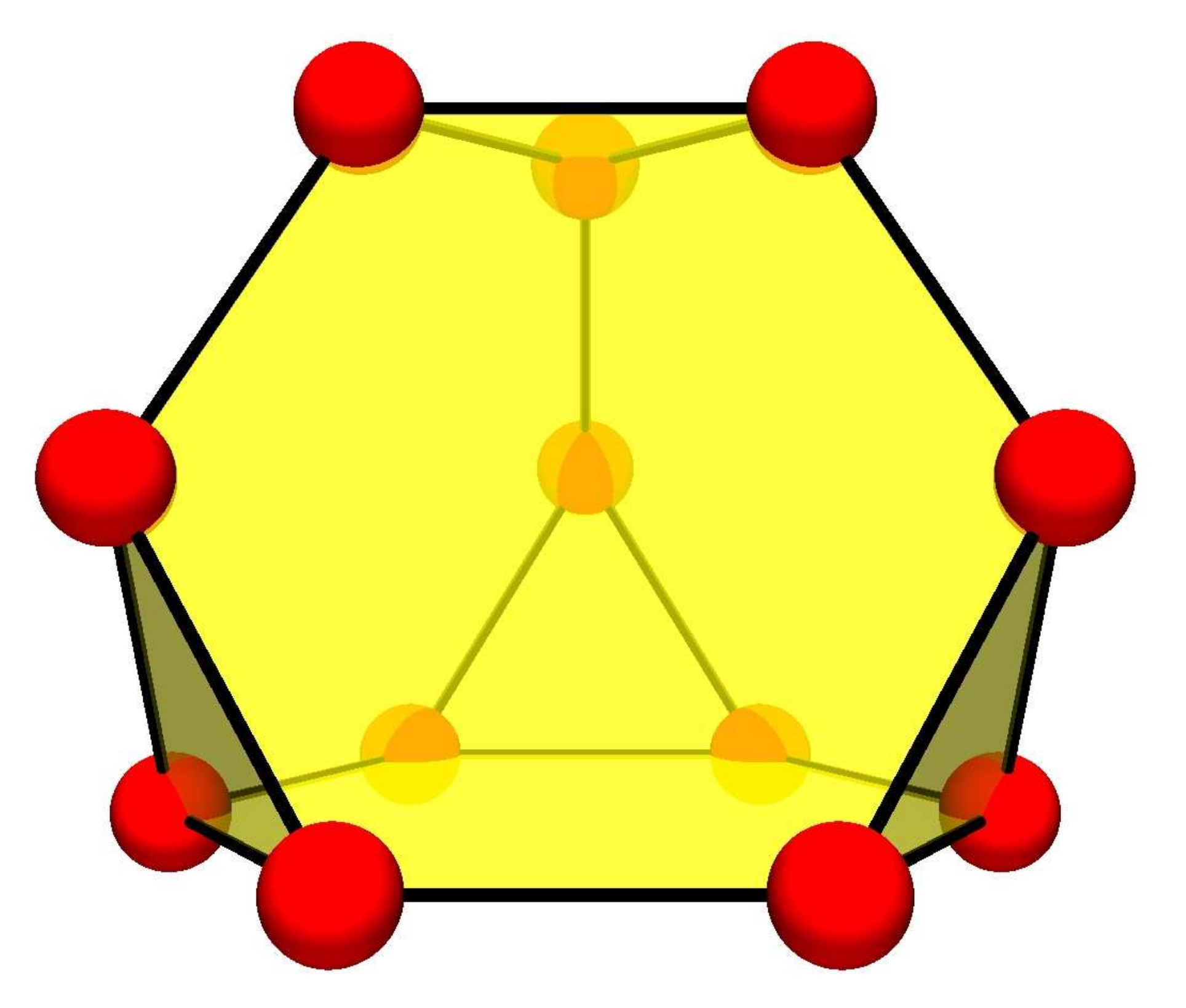}
\caption{The core structure of  \{Gd$_{12}$Mo$_4$\} is a truncated
  tetrahedron. The bullets represent the 12 spin sites
  and the edges correspond to the 18 exchange interactions between
  nearest-neighbor spins. The exchange inside the four triangles
  is named $J_1$, between triangles $J_2$.}
\label{am-fig-1}
\end{figure}

In a recent publication heterometallic cluster complexes
\{Ln$_{12}$Mo$_4$\} featuring a Ln$_{12}$ core that has the
structure of a distorted
truncated tetrahedron, see \figref{am-fig-1}, were reported
\cite{ZZL:CC13}. The 
experimental magnetic studies of the \{Gd$_{12}$Mo$_4$\} were
accompanied on the theoretical side by calculations that
replaced the Gd spin of $s=7/2$ by fictitious spins $s=5/2$
since otherwise the calculation would not have been feasible in
a reasonable time (of several weeks [sic!]). Now, after a few
months, the calculations using the Finite-Temperature Lanczos
Method for $N=12$ spins $s=7/2$ are completed. Before presenting
the results a short reminder of the method shall be given which
in detail is explained elsewhere
\cite{PhysRevB.49.5065,JaP:AP00,ScW:EPJB10}. 

For the evaluation of thermodynamic properties in the canonical
ensemble the exact partition function $Z$ depending on
temperature $T$ and magnetic field $B$ is given by 
\begin{eqnarray}
\label{E-1-1}
Z(T,B)
&=&
\sum_{\nu}\;
\bra{\nu} e^{-\beta \op{H}} \ket{\nu}
\ .
\end{eqnarray}
Here $\{\ket{\nu}\}$ denotes an orthonormal basis of the
respective Hilbert space. Following the ideas of
Refs.~\cite{PhysRevB.49.5065,JaP:AP00} the unknown matrix
elements are approximated as
\begin{eqnarray}
\label{E-1-2}
\bra{\nu} e^{-\beta \op{H}} \ket{\nu}
&\approx&
\sum_{n=1}^{N_L}\;
\braket{\nu}{n(\nu)} e^{-\beta \epsilon_n^{(\nu)}} \braket{n(\nu)}{\nu}
\ ,
\end{eqnarray}
which yields for the partition function
\begin{eqnarray}
\label{E-1-3}
Z(T,B)
&\approx&
\frac{\text{dim}({\mathcal H})}{R}
\sum_{\nu=1}^R\;
\sum_{n=1}^{N_L}\;
e^{-\beta \epsilon_n^{(\nu)}} |\braket{n(\nu)}{\nu}|^2
\ .
\end{eqnarray}
For this procedure $\ket{\nu}$ is taken as the initial vector of a Lanczos
iteration. This iteration consists of $N_L$ Lanczos steps, which
span a respective Krylow space, in which the Hamiltonian is
diagonalized. This yields the $N_L$ Lanczos eigenvectors
$\ket{n(\nu)}$ as well as the associated Lanczos energy
eigenvalues $\epsilon_n^{(\nu)}$. They are enumerated by
$n=1,\dots, N_L$. The number of Lanczos steps $N_L$ is a
parameter of the approximation; $N_L\approx 100$
is usually a good value. 
In addition, the complete and thus very large sum over all
states $\ket{\nu}$ is replaced by a summation over a subset of $R$ random
vectors, where $R$ is the second parameter of the method. For
many cases $R$ can be rather small, e.g. $R\approx 20$, whereas
for other systems convergence is achieved only for $R\approx
100$. 
An observable would then be calculated as
\begin{eqnarray}
\label{E-1-5}
O(T,B)
&\approx&
\frac{1}{Z(T,B)}
\sum_{\Gamma}\;
\frac{\text{dim}({\mathcal H}(\Gamma))}{R_{\Gamma}}
\sum_{\nu=1}^{R_{\Gamma}}\;
\sum_{n=1}^{N_L}\;
e^{-\beta \epsilon_n^{(\nu,\Gamma)}}
\nonumber \\
&&\times
\bra{n(\nu, \Gamma)}\op{O}\ket{\nu, \Gamma}
\braket{\nu, \Gamma}{n(\nu, \Gamma)}
\ .
\end{eqnarray}
Here $\Gamma$ labels the irreducible representations of a
symmetry group that can be used to split the Hilbert space into
subspaces  ${\mathcal H}(\Gamma)$ in order to increase the
accuracy. In
the following calculations we decomposed the Hilbert space
according to the total magnetic quantum number $M$. 

\begin{figure}[ht!]
\centering
\includegraphics*[clip,width=60mm]{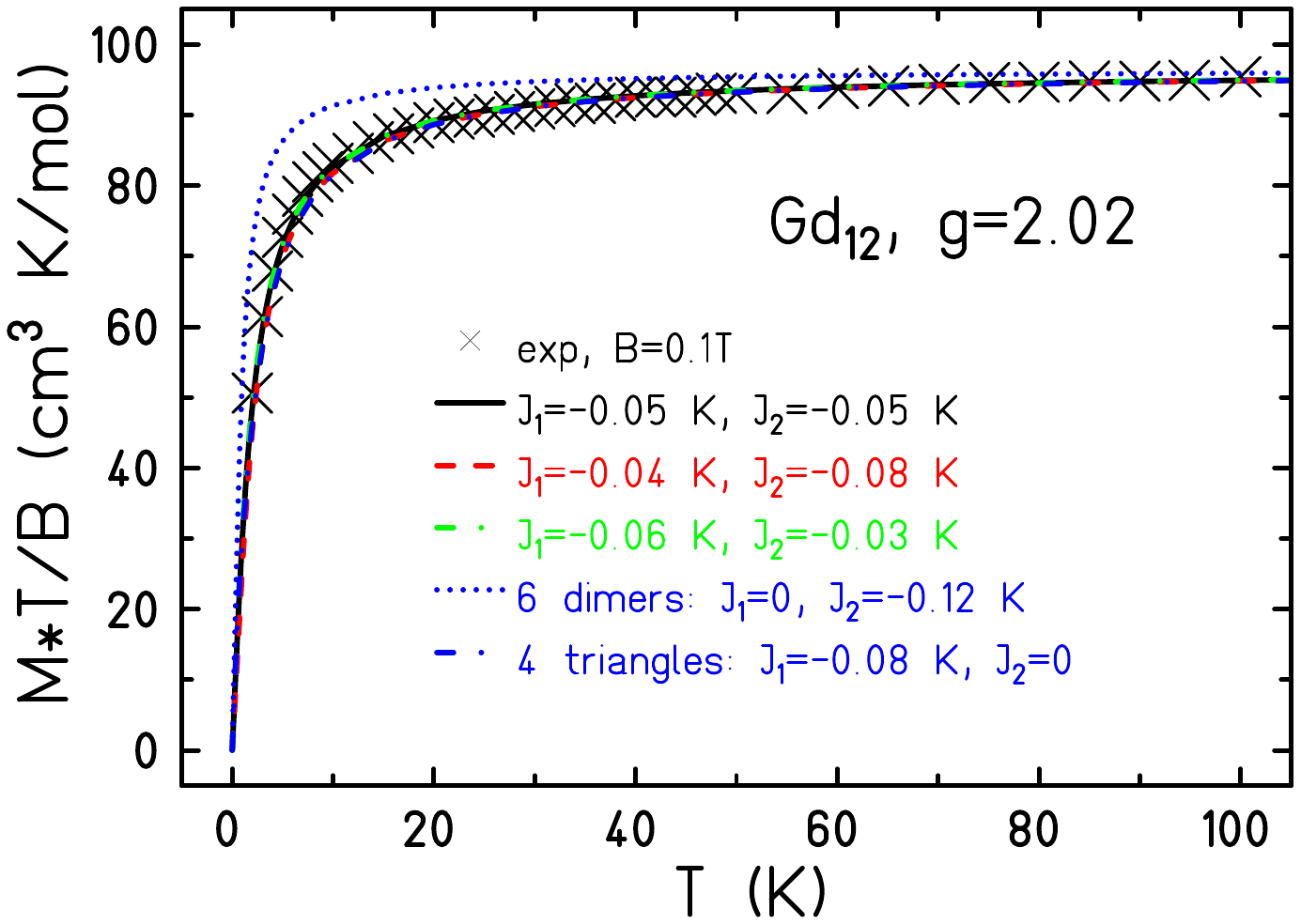}
\includegraphics*[clip,width=60mm]{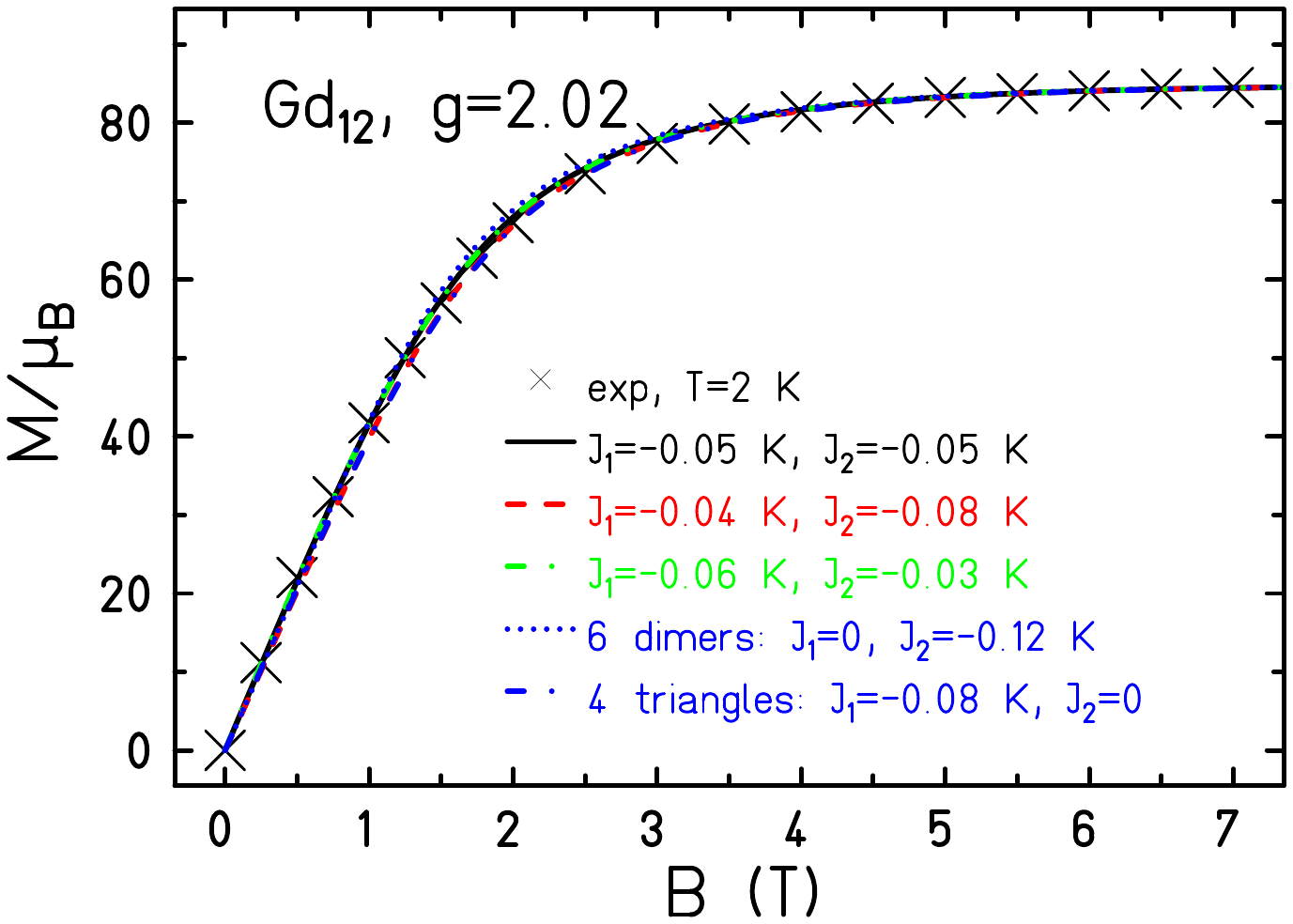}
\caption{Magnetization of \{Gd$_{12}$Mo$_4$\} as function of
  temperature (l.h.s.) as well as of applied magnetic field
  (r.h.s.). The experimental data \cite{ZZL:CC13} are given as
  symbols, the theoretical calculations as curves.}
\label{am-fig-2}
\end{figure}

The magnetization of \{Gd$_{12}$Mo$_4$\} was evaluated for four
different parameter sets. Since the total dimension is a
staggering $(2 s + 1)^N=68,719,476,736$ and even the dimension
of the largest Hilbert subspace with $M=0$ is still
$3,409,213,016$, the
calculations needed about a quarter of a year on a
supercomputer. As \figref{am-fig-2} shows, the exchange
interactions are antiferromagnetic and of the order of
-0.05~K. Since they are so small, the experimental data, taken
from \cite{ZZL:CC13}, is not sufficient to disentangle between
scenarios where the 
interactions $J_1$ between spins within triangles and $J_2$
between triangles are the same or different. A
scenario where only interactions between triangles bind the
spins into dimers can be excluded, but a scenario where the
system would consist of uncoupled triangles cannot be excluded. 

\begin{figure}[ht!]
\centering
\includegraphics*[clip,width=60mm]{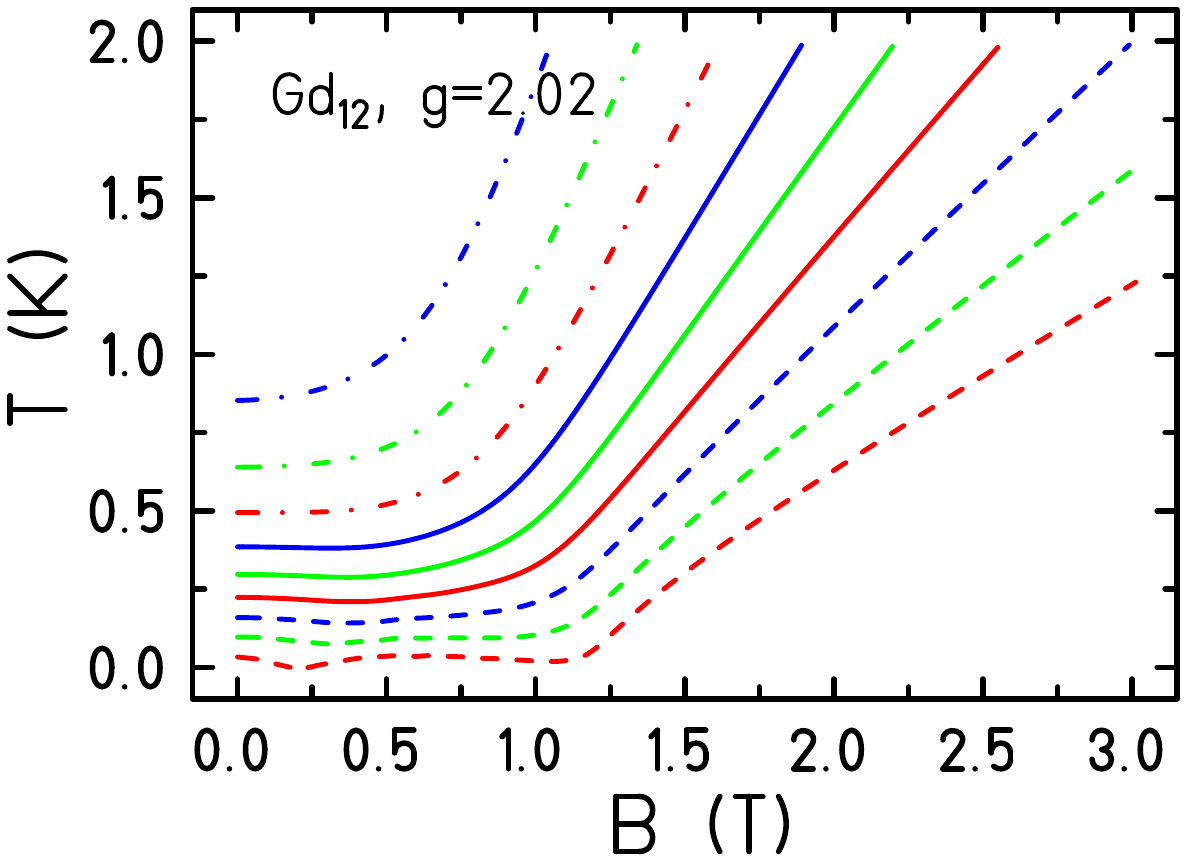}
\includegraphics*[clip,width=60mm]{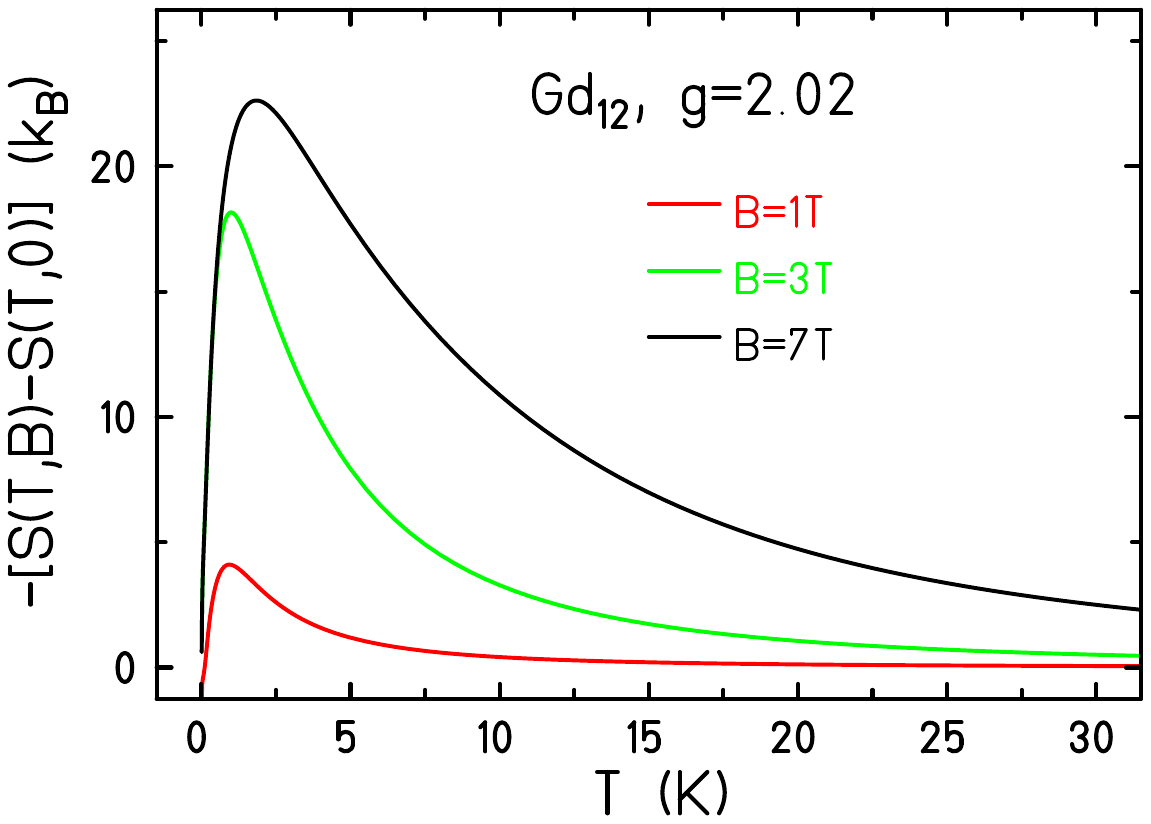}
\caption{Theoretical magnetic isentropes \{Gd$_{12}$Mo$_4$\} as
  function of temperature and field (l.h.s.) and isothermal
  entropy change for various starting fields (r.h.s.). The
  chosen parameters are $J_1=J_2=-0.05$~K.}
\label{am-fig-3}
\end{figure}

Figure \xref{am-fig-3} displays the magnetocaloric behavior for a
coupling scheme with $J_1=J_2=-0.05$~K. The l.h.s. shows a set
of isentropes, i.e. curves which the system would follow when
the magnetic field is reduced in an adiabatic process. The
figure on the r.h.s. shows the isothermal entropy changes for
field sweeps from $B=(1,2,7)$~T, respectively, to $B=0$. The
entropy differences are rather large at low temperature as
expected for a weakly couple gadolinium system.

\section{Quantum Monte Carlo}
\label{sec-3}

Quantum Monte Carlo (QMC) \cite{SaK:PRB91,San:PRB99,San:AIPCP10}
is a very powerful method for non-frustrated, i.e. bipartite
quantum spin systems. For a discussion of frustration see
e.g. \cite{Sch:DT10}. The method can easily deal with up to 100
or more spins. In the field of molecule-based magnetism is was
applied to several spin systems, e.g. homo- and heterometallic
rings \cite{EnL:PRB06,EMP:ACIE08,EMP:PRB09,TMB:CC09} as well as
to a one-dimensional spin tube \cite{ISS:PRL10}. In the latter
publication the heat capacity of a system of $N=100$ spins with
$s=3/2$ was calculated with the help of QMC.

\begin{figure}[ht!]
\centering
\includegraphics*[clip,width=80mm]{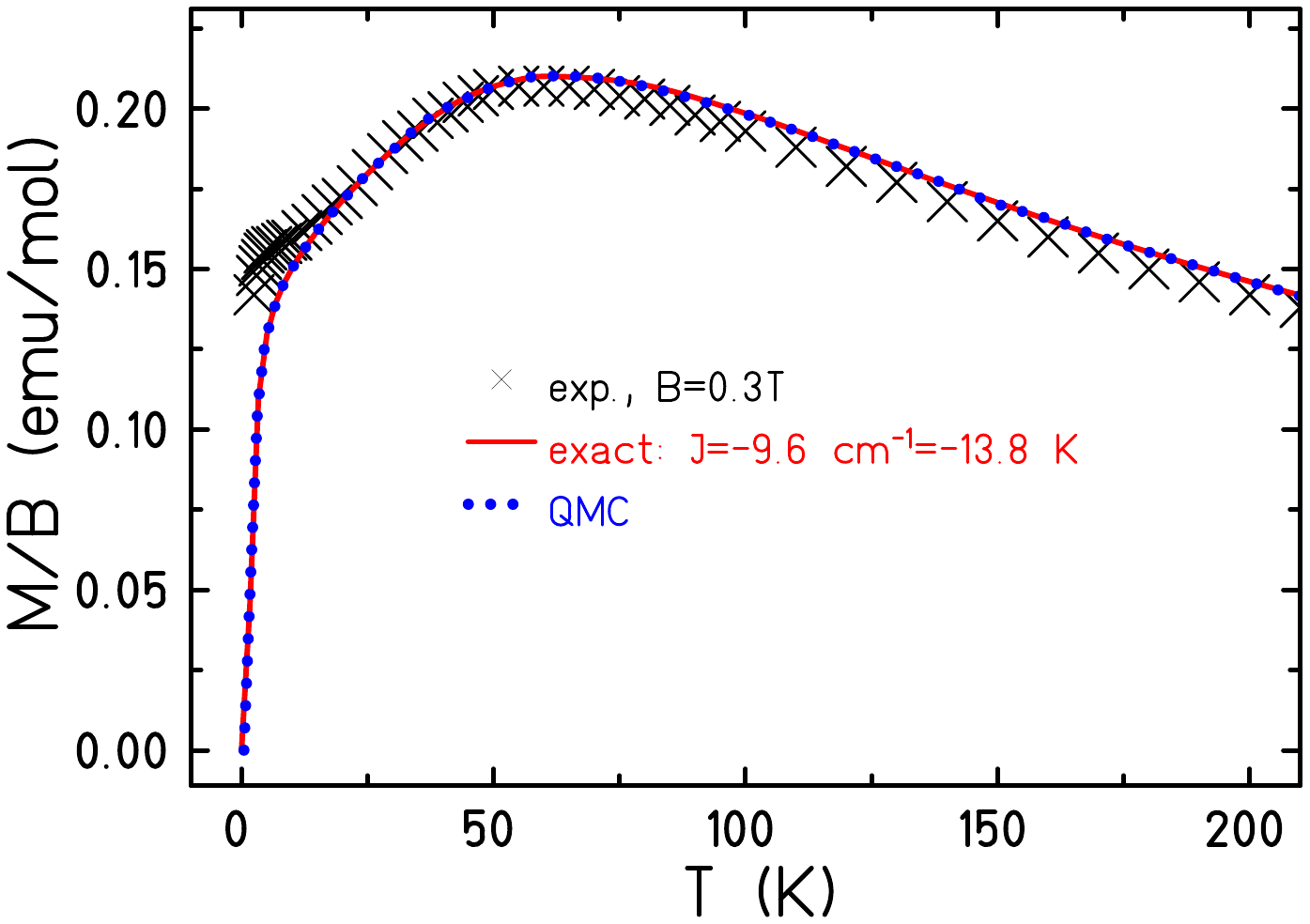}
\caption{Susceptibility of an antiferromagnetically coupled spin
  ring with $N=10$ and $s=5/2$. The exchange parameter
  $J=-9.6$~cm$^{-1}$ as well as 
  the susceptibility data are taken from
  Ref.~\cite{TDP:JACS94}. The solid curve displays the result of
  complete matrix diagonalization \cite{ScS:IRPC10} whereas the dotted
  curve depicts the QMC result.}
\label{am-fig-4}
\end{figure}

Again the idea is to approximate the partition function. This
time the partition function is chopped (sliced) in the sense
that the exponential is written as a product of $m$ exponentials
with exponents divided by $m$ (Trotter-Suzuki decomposition
\cite{Tro:PAMS59,Suz:CMP76,Suz:CMP77}). For $m\rightarrow\infty$ 
the exponential can be written as a product of the exponentials
of the (even non-commuting) parts of the Hamiltonian. One can as
well linearize the exponential for large enough $m$. In any case,
the multi-index sum 
is evaluated in a Monte-Carlo fashion as sketched in the
equations below:
\begin{eqnarray}
\label{E-3-1}
Z(T,B)
&=&
\sum_{\nu}\;
\bra{\nu} e^{-\beta \op{H}} \ket{\nu}
=
\sum_{\nu}\;
\bra{\nu} \left[\exp\left\{-\beta \op{H}/m\right\} \right]^m
\ket{\nu}
\\
&=&
\sum_{\nu, \alpha, \gamma, \dots}\;\bra{\nu} 
\exp\left\{-\beta \op{H}/m\right\} 
\ket{\alpha}\bra{\alpha} 
\exp\left\{-\beta \op{H}/m\right\} 
\ket{\gamma}\bra{\gamma} 
\cdots
\nonumber
\\
&\approx&
\sum_{\nu, \alpha, \gamma, \dots}\;\bra{\nu} \left\{1-\beta
\op{H}/m\right\} \ket{\alpha}\bra{\alpha} \cdots
\nonumber
\ .
\end{eqnarray}
As an example the magnetic susceptibility of the ten-membered
ferric wheel, which was synthesized 18 years ago
\cite{TDP:JACS94}, is presented in \figref{am-fig-4}. The symbols
mark the experimental values \cite{TDP:JACS94}, the solid curve shows the result
of exact diagonalization \cite{ScS:IRPC10}, and the dotted curve shows the result
obtained with ALPS QMC using $10^7$ steps for equilibration and
$10^{10}$ steps for the Monte-Carlo sampling for each
temperature. As one can see the QMC result is indistinguishable from 
the exact one. 

\section{DMRG results}
\label{sec-4}

\begin{figure}[ht!]
\centering
\includegraphics*[clip,width=80mm]{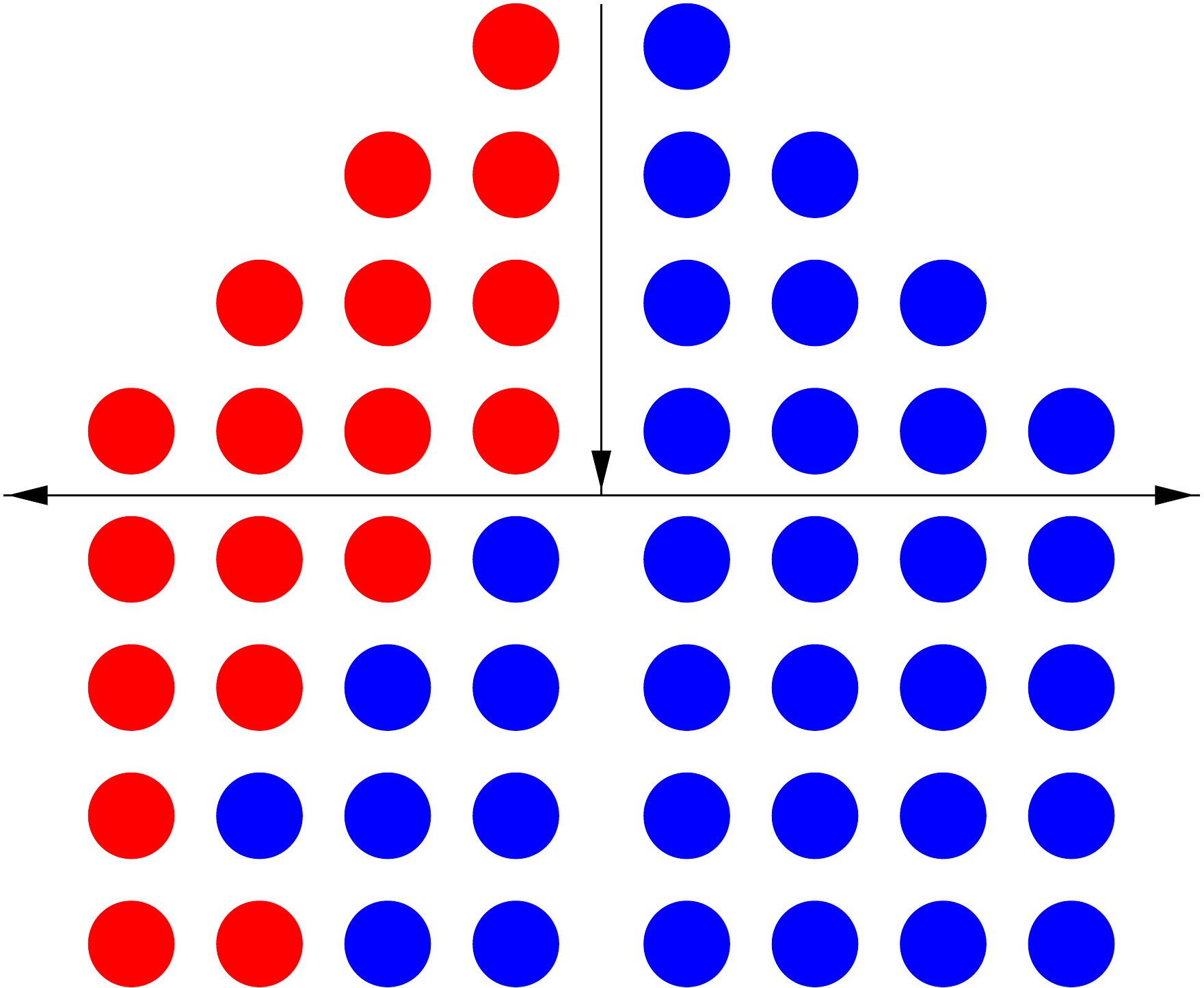}
\caption{Scheme of DMRG: first the system is iteratively
  enlarged up to its full size, then a sweep algorithm improves
  the accuracy further.}
\label{am-fig-5}
\end{figure}

DMRG is a technically rather involved method. One should
however keep in mind that it is a variational method that
constructs trial states in a certain way. Being variational
means that a lower energy for the trial ground state corresponds
to having obtained a better approximation since the energy is
bounded from below by the true value (Ritz's variational
principle). A maybe helpful (mis-) conception of the idea of the
method could be the following: Assume that you want to describe
an eight-membered spin chain as depicted in \figref{am-fig-5}. One
starts with a small subsystem of two spins, diagonalizes the
Hamiltonian and keeps only the lowest $m$ eigenstates. Then one
adds spins sequentially and every time sets up a new basis
built of the old kept states and the states representing the
added spin, diagonalizes the Hamiltonian and again keeps the
lowest $m$ states. This idea is brilliant (and called Numerical
Renormalization Group), except that it does not work in this
naive fashion. Steve White found out, that instead of keeping the
lowest $m$ eigenstates of the Hamiltonian it is much better to
keep $m$ eigenstates of a reduced density matrix in order to iteratively built up the
system and to represent the Hamiltonian \cite{Whi:PRL1992}. 
The density matrix is given by the target state, e.g. the
ground state $\ket{\Psi}$, as $\rho=\ket{\Psi}\bra{\Psi}$ and to
reduce it means to trace over a part of the system, i.e. one of
the colored parts in \figref{am-fig-5}. The
representation can be further improved by running a so-called
sweep algorithm in which the system is subdivided into unequal
blocks for which the Hamiltonian is diagonalized and the density
matrix calculated. 

For the non-expert this seems to be rather obscure, but contrary
to several other methods, DMRG is (1) variational, (2) a
controlled approximation, i.e. with $m\rightarrow\infty$ one
approaches the exact result, and (3) offers accuracy estimators
in the form of the truncated weight or the entanglement
entropy. An extrapolation to the exact result is thus possible
by using these measures. 
Although DMRG is best suited for open one-dimensional chain
systems it can be applied to finite-size clusters, too. The
resulting convergence, which is exponential in $m$ for
one-dimensional chains, is somewhat slower, e.g. like $1/m$ for
a spin cluster such as the icosidodecahedron \cite{ExS:PRB03}. 

Here we would like to demonstrate the power of the method by
showing theoretical magnetization curves for another
ferric wheel, this time an Fe$_{18}$ ring of $N=18$ spins with
$s=5/2$ \cite{KSA:ACIE06}. The dimension of the Hilbert space
for this system is about $10^{14}$ which again renders an exact
treatment impossible. The molecule was investigated by means of
DMRG and DDMRG \cite{UNM:PRB12}, the later results were compared to INS data and
utilized to fix the parameters of the model.

\begin{figure}[ht!]
	\centering
		\includegraphics[width=80mm]{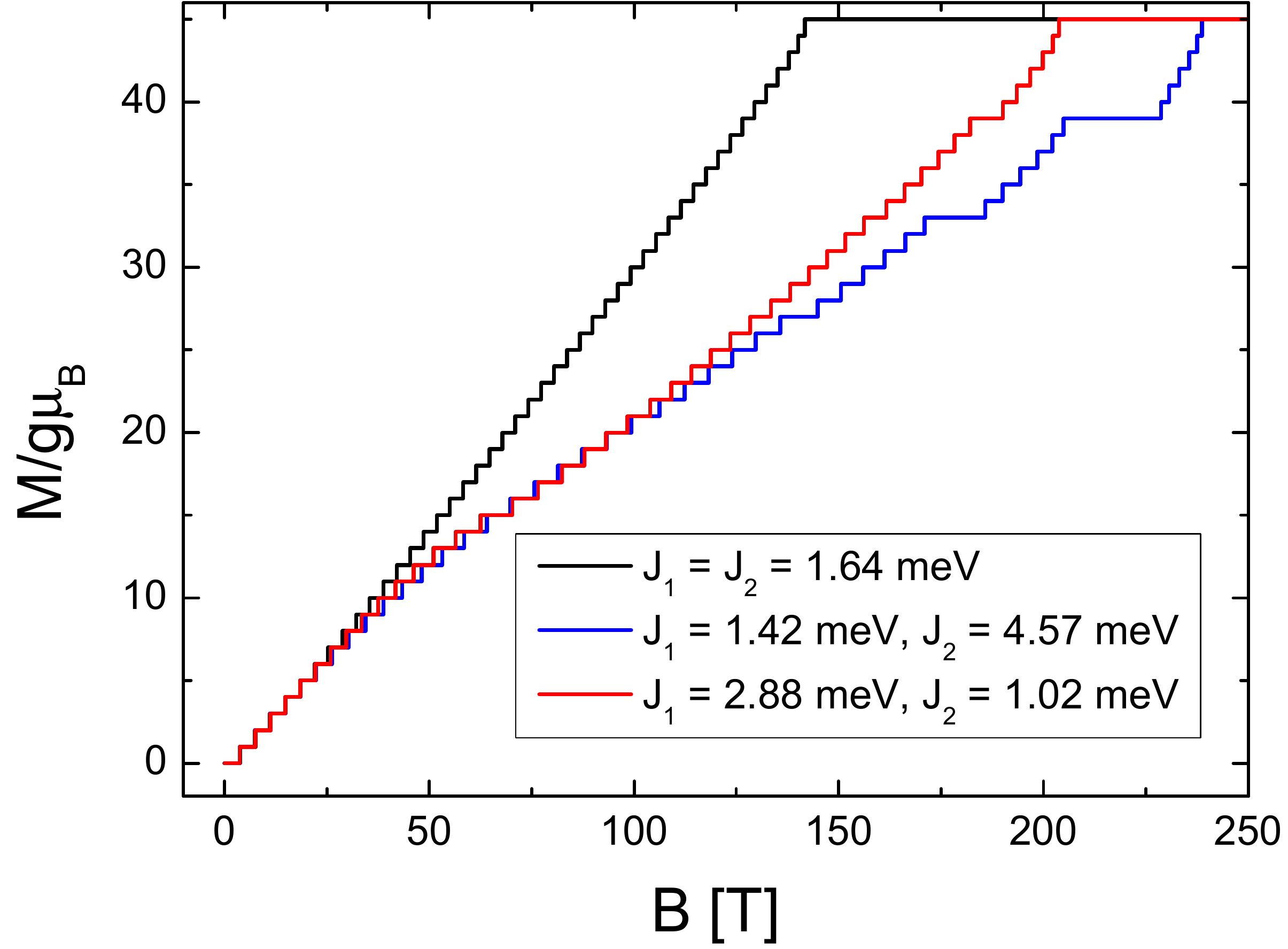}
	\caption{Comparison of the zero-temperature
	magnetization curves for three different parameter sets
	as obtained by standard DMRG calculations for the
	$N=18$, $s=5/2$ spin ring. We have kept up to 3000 density matrix
	eigenstates for the $J_1=J_2$ model and up to 1200 for
	the $J_1\neq J_2$ models. The truncated weights are
	smaller than $10^{-8}$.} 
	\label{am-fig-6}
\end{figure}

In \figref{am-fig-6} we present the $(T=0)$ magnetization curves for the three
parameterizations discussed in \cite{UNM:PRB12}. The results are
very interesting. The magnetization curves are 
virtually identical up to 25~Tesla. The step widths of the
magnetization curve for a $C_{18}$-symmetric ring (single $J$)
are approximately the same for 
all steps (apart from the very last steps), as would also be the
case for the rotational band approximation \cite{ScL:PRB00}. The magnetization
curves for the other two models ($C_6$-symmetric:
$J_1$--$J_1$--$J_2$) deviate from this behavior. Up 
to approximately 100~T, the two models with $J_1\neq J_2$ give
very similar magnetization curves and considerable differences
appear only at even higher fields. The magnetization curve for
$J_1=1.42$~meV and $J_2=4.57$~meV shows two magnetization
plateaus at higher fields. Plateaus in zero-temperature
magnetization curves usually emerge in geometrically frustrated
spin systems \cite{HSR:JP04}. This system is,
however, not geometrically frustrated so that the emergence of a
plateau is an interesting effect. 

\begin{figure}[ht!]
	\centering
		\includegraphics[width=1.0\textwidth]{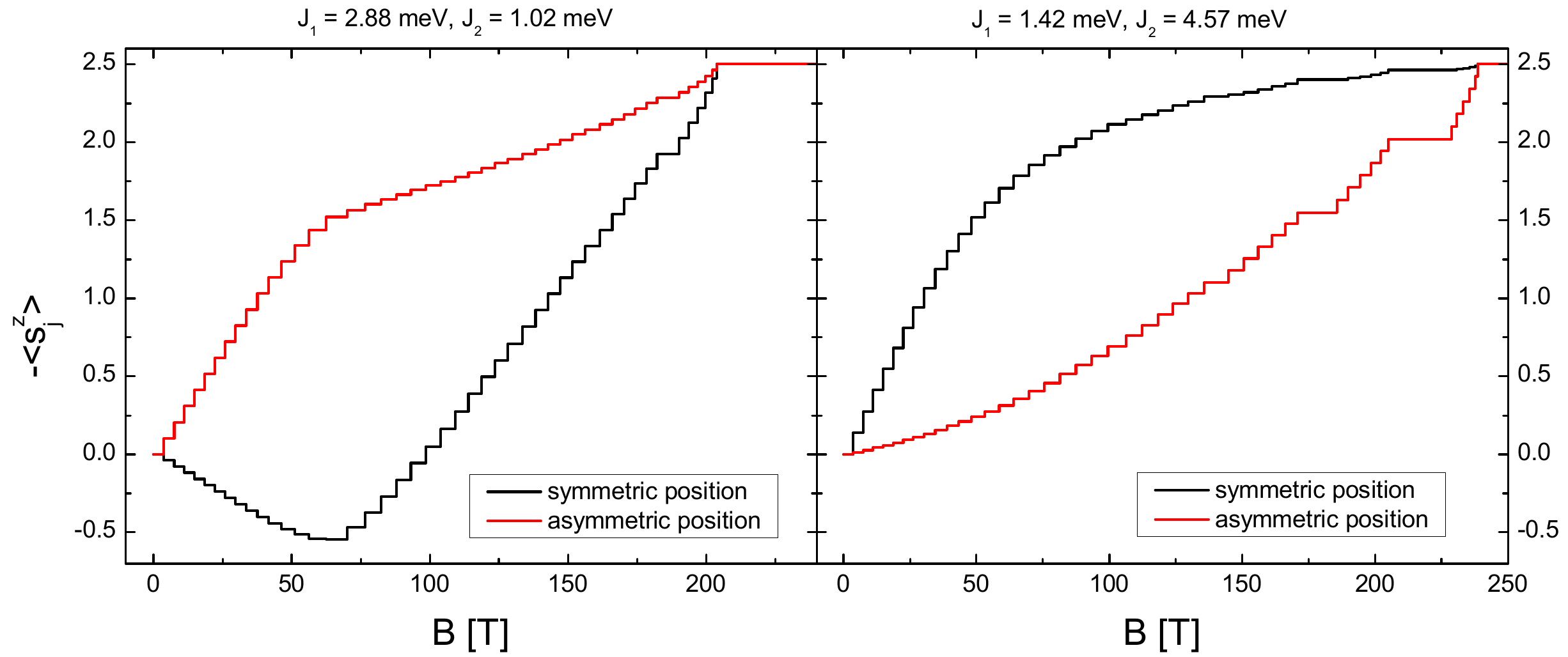}
	\caption{Dependence of the $z$-component of the local
	magnetization (which is proportional to $-\langle
	s^z_j\rangle$) for two different parameter sets as a
	function of an external magnetic field
	$\vec{B}=B\vec{e}_z$. The results were obtained using
	standard DMRG and $T=0$. The result for the
	uniform model is not shown because for this model the
	local moments are simply proportional to the
	magnetization curve presented in
	Fig.~\ref{am-fig-6}.
	``Symmetric position'' denotes the local magnetization
	for a spin between two $J_1$ couplings. Accordingly,
	``asymmetric position'' denotes a spin position between
	a $J_1$ and a $J_2$ coupling.} 
	\label{am-fig-7}
\end{figure}

We have also calculated the local magnetizations for the two
models with $J_1\neq J_2$. Local moments can,
e.g., be probed with NMR \cite{MFK:PRL06,FNK:PRB07} or XMCD
\cite{PKT:IC10}. The local 
magnetizations for the $J_1=J_2$ model would simply be
proportional to the total magnetization curve. The results are
shown in \figref{am-fig-7}. The calculation demonstrates that the
local magnetization of an interacting spin system can deviate
substantially from the average and even point into the opposite
direction.

\section*{Acknowledgment}

This work was supported by the German Science Foundation (DFG)
through the research group 945. Computing time at the Leibniz
Computing Center in Garching is also gratefully
acknowledged. Last but not least we like to thank the State of
North Rhine-Westphalia and the DFG for financing our local SMP
supercomputer as well as the companies BULL and ScaleMP for
their support.


\begin{thebibliography}{59}
\providecommand{\natexlab}[1]{#1}
\providecommand{\url}[1]{\texttt{#1}}
\providecommand{\urlprefix}{URL }
\expandafter\ifx\csname urlstyle\endcsname\relax
  \providecommand{\doi}[1]{doi:\discretionary{}{}{}#1}\else
  \providecommand{\doi}[1]{doi:\discretionary{}{}{}\begingroup
  \urlstyle{rm}\url{#1}\endgroup}\fi
\providecommand{\bibinfo}[2]{#2}

\bibitem[{Gatteschi and Pardi(1993)}]{GaP:GCI93}
\bibinfo{author}{D.~Gatteschi}, \bibinfo{author}{L.~Pardi},
  \bibinfo{title}{Magnetic-properties of high-nuclearity spin clusters - a fast
  and efficient procedure for the calculation of the energy-levels},
  \bibinfo{journal}{Gazz. Chim. Ital.} \bibinfo{volume}{123}
  (\bibinfo{year}{1993}) \bibinfo{pages}{231--240}.

\bibitem[{Borras-Almenar et~al.(1999)Borras-Almenar, Clemente-Juan, Coronado,
  and Tsukerblat}]{BCC:IC99}
\bibinfo{author}{J.~J. Borras-Almenar}, \bibinfo{author}{J.~M. Clemente-Juan},
  \bibinfo{author}{E.~Coronado}, \bibinfo{author}{B.~S. Tsukerblat},
  \bibinfo{title}{High-nuclearity magnetic clusters: Generalized spin
  Hamiltonian and its use for the calculation of the energy levels, bulk
  magnetic properties, and inelastic neutron scattering spectra},
  \bibinfo{journal}{Inorg. Chem.} \bibinfo{volume}{38} (\bibinfo{year}{1999})
  \bibinfo{pages}{6081--6088}.

\bibitem[{Bencini and Gatteschi(1990)}]{BeG:EPR}
\bibinfo{author}{A.~Bencini}, \bibinfo{author}{D.~Gatteschi},
  \bibinfo{title}{Electron paramagnetic resonance of exchange coupled systems},
  \bibinfo{publisher}{Springer}, \bibinfo{address}{Berlin, Heidelberg},
  \bibinfo{year}{1990}.

\bibitem[{Tsukerblat(2006)}]{Tsu:group_theory}
\bibinfo{author}{B.~S. Tsukerblat}, \bibinfo{title}{Group theory in chemistry
  and spectroscopy: a simple guide to advanced usage},
  \bibinfo{publisher}{Dover Publications}, \bibinfo{address}{Mineola, New
  York}, \bibinfo{edition}{2nd} edn., \bibinfo{year}{2006}.

\bibitem[{Tsukerblat(2008)}]{Tsu:ICA08}
\bibinfo{author}{B.~Tsukerblat}, \bibinfo{title}{Group-theoretical approaches
  in molecular magnetism: Metal clusters}, \bibinfo{journal}{Inorg. Chim. Acta}
  \bibinfo{volume}{361} (\bibinfo{year}{2008}) \bibinfo{pages}{3746--3760}.

\bibitem[{Boyarchenkov et~al.(2007)Boyarchenkov, Bostrem, and
  Ovchinnikov}]{BBO:PRB07}
\bibinfo{author}{A.~S. Boyarchenkov}, \bibinfo{author}{I.~G. Bostrem},
  \bibinfo{author}{A.~S. Ovchinnikov}, \bibinfo{title}{Quantum magnetization
  plateau and sign change of the magnetocaloric effect in a ferrimagnetic spin
  chain.}, \bibinfo{journal}{Phys. Rev. B} \bibinfo{volume}{76}
  (\bibinfo{year}{2007}) \bibinfo{pages}{224410},
  \doi{\bibinfo{doi}{10.1103/PhysRevB.76.224410}}.

\bibitem[{Delfs et~al.(1993)Delfs, Gatteschi, Pardi, Sessoli, Wieghardt, and
  Hanke}]{DGP:IC93}
\bibinfo{author}{C.~Delfs}, \bibinfo{author}{D.~Gatteschi},
  \bibinfo{author}{L.~Pardi}, \bibinfo{author}{R.~Sessoli},
  \bibinfo{author}{K.~Wieghardt}, \bibinfo{author}{D.~Hanke},
  \bibinfo{title}{Magnetic properties of an octanuclear iron(III) cation},
  \bibinfo{journal}{Inorg. Chem.} \bibinfo{volume}{32} (\bibinfo{year}{1993})
  \bibinfo{pages}{3099--3103}, \doi{\bibinfo{doi}{10.1021/ic00066a022}}.

\bibitem[{Waldmann(2000)}]{Wal:PRB00}
\bibinfo{author}{O.~Waldmann}, \bibinfo{title}{Symmetry and energy spectrum of
  high-nuclearity spin clusters}, \bibinfo{journal}{Phys. Rev. B}
  \bibinfo{volume}{61} (\bibinfo{year}{2000}) \bibinfo{pages}{6138}.

\bibitem[{Bostrem et~al.(2006)Bostrem, Ovchinnikov, and Sinitsyn}]{BOS:TMP06}
\bibinfo{author}{I.~G. Bostrem}, \bibinfo{author}{A.~S. Ovchinnikov},
  \bibinfo{author}{V.~E. Sinitsyn}, \bibinfo{title}{The method of exact
  diagonalization preserving the total spin and taking the point symmetry of
  the two-dimensional isotropic Heisenberg magnet into account},
  \bibinfo{journal}{Theor. Math. Phys.} \bibinfo{volume}{149}
  (\bibinfo{year}{2006}) \bibinfo{pages}{1527--1544}.

\bibitem[{Sinitsyn et~al.(2007)Sinitsyn, Bostrem, and Ovchinnikov}]{SBO:JPA07}
\bibinfo{author}{V.~E. Sinitsyn}, \bibinfo{author}{I.~G. Bostrem},
  \bibinfo{author}{A.~S. Ovchinnikov}, \bibinfo{title}{Symmetry adapted
  finite-cluster solver for quantum Heisenberg model in two dimensions: a
  real-space renormalization approach}, \bibinfo{journal}{J. Phys. A-Math.
  Theor.} \bibinfo{volume}{40} (\bibinfo{year}{2007})
  \bibinfo{pages}{645--668}.

\bibitem[{Schnalle and Schnack(2009)}]{ScS:PRB09}
\bibinfo{author}{R.~Schnalle}, \bibinfo{author}{J.~Schnack},
  \bibinfo{title}{Numerically exact and approximate determination of energy
  eigenvalues for antiferromagnetic molecules using irreducible tensor
  operators and general point-group symmetries}, \bibinfo{journal}{Phys. Rev.
  B} \bibinfo{volume}{79} (\bibinfo{year}{2009}) \bibinfo{pages}{104419},
  \doi{\bibinfo{doi}{10.1103/PhysRevB.79.104419}}.

\bibitem[{Schnack and Schnalle(2009)}]{ScS:P09}
\bibinfo{author}{J.~Schnack}, \bibinfo{author}{R.~Schnalle},
  \bibinfo{title}{Frustration effects in antiferromagnetic molecules: the
  cuboctahedron}, \bibinfo{journal}{Polyhedron} \bibinfo{volume}{28}
  (\bibinfo{year}{2009}) \bibinfo{pages}{1620--1623},
  \doi{\bibinfo{doi}{10.1016/j.poly.2008.10.017}}.

\bibitem[{Schnalle and Schnack(2010)}]{ScS:IRPC10}
\bibinfo{author}{R.~Schnalle}, \bibinfo{author}{J.~Schnack},
  \bibinfo{title}{Calculating the energy spectra of magnetic molecules:
  application of real- and spin-space symmetries}, \bibinfo{journal}{Int. Rev.
  Phys. Chem.} \bibinfo{volume}{29} (\bibinfo{year}{2010})
  \bibinfo{pages}{403--452}, \doi{\bibinfo{doi}{10.1080/0144235X.2010.485755}}.

\bibitem[{Jaklic and Prelovsek(1994)}]{PhysRevB.49.5065}
\bibinfo{author}{J.~Jaklic}, \bibinfo{author}{P.~Prelovsek},
  \bibinfo{title}{Lanczos method for the calculation of finite-temperature
  quantities in correlated systems}, \bibinfo{journal}{Phys. Rev. B}
  \bibinfo{volume}{49} (\bibinfo{year}{1994}) \bibinfo{pages}{5065--5068},
  \doi{\bibinfo{doi}{10.1103/PhysRevB.49.5065}}.

\bibitem[{Jaklic and Prelovsek(2000)}]{JaP:AP00}
\bibinfo{author}{J.~Jaklic}, \bibinfo{author}{P.~Prelovsek},
  \bibinfo{title}{Finite-temperature properties of doped antiferromagnets},
  \bibinfo{journal}{Adv. Phys.} \bibinfo{volume}{49} (\bibinfo{year}{2000})
  \bibinfo{pages}{1--92}.

\bibitem[{Schnack and Wendland(2010)}]{ScW:EPJB10}
\bibinfo{author}{J.~Schnack}, \bibinfo{author}{O.~Wendland},
  \bibinfo{title}{Properties of highly frustrated magnetic molecules studied by
  the finite-temperature Lanczos method}, \bibinfo{journal}{Eur. Phys. J. B}
  \bibinfo{volume}{78} (\bibinfo{year}{2010}) \bibinfo{pages}{535--541},
  \doi{\bibinfo{doi}{10.1007/BF01609348}}.

\bibitem[{Graham et~al.(2011)Graham, Douglas, Mathieson, Moggach, Schnack, and
  Murrie}]{GDM:DT11}
\bibinfo{author}{K.~Graham}, \bibinfo{author}{F.~J. Douglas},
  \bibinfo{author}{J.~S. Mathieson}, \bibinfo{author}{S.~A. Moggach},
  \bibinfo{author}{J.~Schnack}, \bibinfo{author}{M.~Murrie},
  \bibinfo{title}{Cubic assembly of a geometrically frustrated {Fe12} spin
  cluster}, \bibinfo{journal}{Dalton Trans.}  (\bibinfo{year}{2011})
  \bibinfo{pages}{--}\doi{\bibinfo{doi}{10.1039/C1DT10910C}},
  \urlprefix\url{http://dx.doi.org/10.1039/C1DT10910C}.

\bibitem[{Hooper et~al.(2012)Hooper, Schnack, Piligkos, Evangelisti, and
  Brechin}]{HSP:ACIE12}
\bibinfo{author}{T.~N. Hooper}, \bibinfo{author}{J.~Schnack},
  \bibinfo{author}{S.~Piligkos}, \bibinfo{author}{M.~Evangelisti},
  \bibinfo{author}{E.~K. Brechin}, \bibinfo{title}{The Importance of Being
  Exchanged:
  [Gd${}^{\text{III}}_{4}$M${}^{\text{II}}_{8}$(OH)$_8$(L)$_8$(O$_2$CR)$_8$]$^%
{4+}$ Clusters for Magnetic Refrigeration}, \bibinfo{journal}{Angew. Chem. Int.
  Ed.} \bibinfo{volume}{51} (\bibinfo{year}{2012}) \bibinfo{pages}{4633--4636},
  \doi{\bibinfo{doi}{10.1002/anie.201200072}}.

\bibitem[{Garlatti(2012)}]{Garlatti:PC}
\bibinfo{author}{E.~Garlatti}, \bibinfo{note}{private communication},
  \bibinfo{year}{2012}.

\bibitem[{Hutchinson(1989)}]{Hut:CSSC89}
\bibinfo{author}{M.~Hutchinson}, \bibinfo{title}{A Stochastic Estimator of the
  Trace of the Influence Matrix for Laplacian Smoothing Splines},
  \bibinfo{journal}{Communications in Statistics - Simulation and Computation}
  \bibinfo{volume}{18}~(\bibinfo{number}{3}) (\bibinfo{year}{1989})
  \bibinfo{pages}{1059--1076}, \doi{\bibinfo{doi}{10.1080/03610918908812806}}.

\bibitem[{Manthe and Huarte-Larranaga(2001)}]{MHL:CPL01}
\bibinfo{author}{U.~Manthe}, \bibinfo{author}{F.~Huarte-Larranaga},
  \bibinfo{title}{Partition functions for reaction rate calculations:
  statistical sampling and MCTDH propagation}, \bibinfo{journal}{Chem. Phys.
  Lett.} \bibinfo{volume}{349} (\bibinfo{year}{2001}) \bibinfo{pages}{321 --
  328}, \doi{\bibinfo{doi}{10.1016/S0009-2614(01)01207-6}}.

\bibitem[{Huarte-Larranaga and Manthe(2002)}]{HLM:JCP02}
\bibinfo{author}{F.~Huarte-Larranaga}, \bibinfo{author}{U.~Manthe},
  \bibinfo{title}{Vibrational excitation in the transition state: The CH$_4$ +
  H $\rightarrow$ CH$_3$ + H$_2$ reaction rate constant in an extended
  temperature interval}, \bibinfo{journal}{J. Chem. Phys.}
  \bibinfo{volume}{116} (\bibinfo{year}{2002}) \bibinfo{pages}{2863--2869},
  \doi{\bibinfo{doi}{10.1063/1.1436307}}.

\bibitem[{White(1992)}]{Whi:PRL1992}
\bibinfo{author}{S.~R. White}, \bibinfo{title}{Density matrix formulation for
  quantum renormalization groups}, \bibinfo{journal}{Phys. Rev. Lett.}
  \bibinfo{volume}{69}~(\bibinfo{number}{19}) (\bibinfo{year}{1992})
  \bibinfo{pages}{2863--2866},
  \doi{\bibinfo{doi}{10.1103/PhysRevLett.69.2863}}.

\bibitem[{White(1993)}]{Whi:PRB93}
\bibinfo{author}{S.~R. White}, \bibinfo{title}{Density-matrix algorithms for
  quantum renormalization groups}, \bibinfo{journal}{Phys. Rev. B}
  \bibinfo{volume}{48} (\bibinfo{year}{1993}) \bibinfo{pages}{10345}.

\bibitem[{White and Huse(1993)}]{WhD:PRB93B}
\bibinfo{author}{S.~R. White}, \bibinfo{author}{D.~Huse},
  \bibinfo{title}{Numerical renormalization-group study of low-lying
  eigenstates of the antiferromagnetic $s=1$ Heisenberg chain},
  \bibinfo{journal}{Phys. Rev. B} \bibinfo{volume}{48} (\bibinfo{year}{1993})
  \bibinfo{pages}{3844}.

\bibitem[{Schollw\"ock(2005)}]{Sch:RMP05}
\bibinfo{author}{U.~Schollw\"ock}, \bibinfo{title}{The density-matrix
  renormalization group}, \bibinfo{journal}{Rev. Mod. Phys.}
  \bibinfo{volume}{77} (\bibinfo{year}{2005}) \bibinfo{pages}{259--315}.

\bibitem[{Exler and Schnack(2003)}]{ExS:PRB03}
\bibinfo{author}{M.~Exler}, \bibinfo{author}{J.~Schnack},
  \bibinfo{title}{Evaluation of the low-lying energy spectrum of magnetic
  {K}eplerate molecules using the density-matrix renormalization group
  technique}, \bibinfo{journal}{Phys. Rev. B} \bibinfo{volume}{67}
  (\bibinfo{year}{2003}) \bibinfo{pages}{094440},
  \doi{\bibinfo{doi}{10.1103/PhysRevB.67.094440}}.

\bibitem[{Ummethum et~al.(2013)Ummethum, Schnack, and Laeuchli}]{USL:JMMM13}
\bibinfo{author}{J.~Ummethum}, \bibinfo{author}{J.~Schnack},
  \bibinfo{author}{A.~Laeuchli}, \bibinfo{title}{Large-scale numerical
  investigations of the antiferromagnetic Heisenberg icosidodecahedron},
  \bibinfo{journal}{J. Magn. Magn. Mater.} \bibinfo{volume}{327}
  (\bibinfo{year}{2013}) \bibinfo{pages}{103 -- 109},
  \doi{\bibinfo{doi}{10.1016/j.jmmm.2012.09.037}}.

\bibitem[{M\"uller et~al.(1999)M\"uller, Sarkar, Shah, B\"ogge, Schmidtmann,
  Sarkar, K\"ogerler, Hauptfleisch, Trautwein, and Sch\"unemann}]{MSS:ACIE99}
\bibinfo{author}{A.~M\"uller}, \bibinfo{author}{S.~Sarkar},
  \bibinfo{author}{S.~Q.~N. Shah}, \bibinfo{author}{H.~B\"ogge},
  \bibinfo{author}{M.~Schmidtmann}, \bibinfo{author}{S.~Sarkar},
  \bibinfo{author}{P.~K\"ogerler}, \bibinfo{author}{B.~Hauptfleisch},
  \bibinfo{author}{A.~Trautwein}, \bibinfo{author}{V.~Sch\"unemann},
  \bibinfo{title}{Archimedean synthesis and magic numbers: ``Sizing" giant
  molybdenum-oxide-based molecular spheres of the Keplerate type},
  \bibinfo{journal}{Angew. Chem. Int. Ed.} \bibinfo{volume}{38}
  (\bibinfo{year}{1999}) \bibinfo{pages}{3238}.

\bibitem[{M\"uller et~al.(2001)M\"uller, Luban, Schr\"oder, Modler, K\"ogerler,
  Axenovich, Schnack, Canfield, Bud'ko, and Harrison}]{MLS:CPC01}
\bibinfo{author}{A.~M\"uller}, \bibinfo{author}{M.~Luban},
  \bibinfo{author}{C.~Schr\"oder}, \bibinfo{author}{R.~Modler},
  \bibinfo{author}{P.~K\"ogerler}, \bibinfo{author}{M.~Axenovich},
  \bibinfo{author}{J.~Schnack}, \bibinfo{author}{P.~C. Canfield},
  \bibinfo{author}{S.~Bud'ko}, \bibinfo{author}{N.~Harrison},
  \bibinfo{title}{Classical and quantum magnetism in giant Keplerate magnetic
  molecules}, \bibinfo{journal}{Chem. Phys. Chem.} \bibinfo{volume}{2}
  (\bibinfo{year}{2001}) \bibinfo{pages}{517},
  \doi{\bibinfo{doi}{10.1002/1439-7641(20010917)2:8/9<517::AID-CPHC517>3.0.CO;%
2-1}}.

\bibitem[{K\"uhner and White(1999)}]{KuW:PRB99}
\bibinfo{author}{T.~D. K\"uhner}, \bibinfo{author}{S.~R. White},
  \bibinfo{title}{Dynamical correlation functions using the density matrix
  renormalization group}, \bibinfo{journal}{Phys. Rev. B} \bibinfo{volume}{60}
  (\bibinfo{year}{1999}) \bibinfo{pages}{335--343},
  \doi{\bibinfo{doi}{10.1103/PhysRevB.60.335}}.

\bibitem[{Jeckelmann(2002)}]{Jec:PRB02}
\bibinfo{author}{E.~Jeckelmann}, \bibinfo{title}{Dynamical density-matrix
  renormalization-group method}, \bibinfo{journal}{Phys. Rev. B}
  \bibinfo{volume}{66} (\bibinfo{year}{2002}) \bibinfo{pages}{045114},
  \doi{\bibinfo{doi}{10.1103/PhysRevB.66.045114}}.

\bibitem[{Ummethum et~al.(2012)Ummethum, Nehrkorn, Mukherjee, Ivanov, Stuiber,
  Str\"assle, Tregenna-Piggott, Mutka, Christou, Waldmann, and
  Schnack}]{UNM:PRB12}
\bibinfo{author}{J.~Ummethum}, \bibinfo{author}{J.~Nehrkorn},
  \bibinfo{author}{S.~Mukherjee}, \bibinfo{author}{N.~B. Ivanov},
  \bibinfo{author}{S.~Stuiber}, \bibinfo{author}{T.~Str\"assle},
  \bibinfo{author}{P.~L.~W. Tregenna-Piggott}, \bibinfo{author}{H.~Mutka},
  \bibinfo{author}{G.~Christou}, \bibinfo{author}{O.~Waldmann},
  \bibinfo{author}{J.~Schnack}, \bibinfo{title}{Discrete antiferromagnetic
  spin-wave excitations in the giant ferric wheel Fe${}_{18}$},
  \bibinfo{journal}{Phys. Rev. B} \bibinfo{volume}{86} (\bibinfo{year}{2012})
  \bibinfo{pages}{104403}, \doi{\bibinfo{doi}{10.1103/PhysRevB.86.104403}}.

\bibitem[{Sandvik and Kurkij\"arvi(1991)}]{SaK:PRB91}
\bibinfo{author}{A.~W. Sandvik}, \bibinfo{author}{J.~Kurkij\"arvi},
  \bibinfo{title}{Quantum Monte Carlo simulation method for spin systems},
  \bibinfo{journal}{Phys. Rev. B} \bibinfo{volume}{43} (\bibinfo{year}{1991})
  \bibinfo{pages}{5950--5961}, \doi{\bibinfo{doi}{10.1103/PhysRevB.43.5950}}.

\bibitem[{Sandvik(1999)}]{San:PRB99}
\bibinfo{author}{A.~W. Sandvik}, \bibinfo{title}{Stochastic series expansion
  method with operator-loop update}, \bibinfo{journal}{Phys. Rev. B}
  \bibinfo{volume}{59} (\bibinfo{year}{1999}) \bibinfo{pages}{R14157--R14160},
  \doi{\bibinfo{doi}{10.1103/PhysRevB.59.R14157}}.

\bibitem[{Sandvik(2010)}]{San:AIPCP10}
\bibinfo{author}{A.~W. Sandvik}, \bibinfo{title}{Computational Studies of
  Quantum Spin Systems}, \bibinfo{journal}{AIP Conf. Proc.}
  \bibinfo{volume}{1297} (\bibinfo{year}{2010}) \bibinfo{pages}{135--338},
  \doi{\bibinfo{doi}{10.1063/1.3518900}}.

\bibitem[{Henelius and Sandvik(2000)}]{HeS:PRB00}
\bibinfo{author}{P.~Henelius}, \bibinfo{author}{A.~W. Sandvik},
  \bibinfo{title}{Sign problem in Monte Carlo simulations of frustrated quantum
  spin systems}, \bibinfo{journal}{Phys. Rev. B} \bibinfo{volume}{62}
  (\bibinfo{year}{2000}) \bibinfo{pages}{1102--1113}.

\bibitem[{Engelhardt and Luban(2006)}]{EnL:PRB06}
\bibinfo{author}{L.~Engelhardt}, \bibinfo{author}{M.~Luban},
  \bibinfo{title}{Low temperature magnetization and the excitation spectrum of
  antiferromagnetic Heisenberg spin rings}, \bibinfo{journal}{Phys. Rev. B}
  \bibinfo{volume}{73} (\bibinfo{year}{2006}) \bibinfo{pages}{054430}.

\bibitem[{Engelhardt et~al.(2008)Engelhardt, Muryn, Pritchard, Timco, Tuna, and
  Winpenny}]{EMP:ACIE08}
\bibinfo{author}{L.~P. Engelhardt}, \bibinfo{author}{C.~A. Muryn},
  \bibinfo{author}{R.~G. Pritchard}, \bibinfo{author}{G.~A. Timco},
  \bibinfo{author}{F.~Tuna}, \bibinfo{author}{R.~E.~P. Winpenny},
  \bibinfo{title}{Octa-, deca-, trideca-, and tetradecanuclear heterometallic
  cyclic chromium-copper cages}, \bibinfo{journal}{Angew. Chem. Int. Edit.}
  \bibinfo{volume}{47} (\bibinfo{year}{2008}) \bibinfo{pages}{924--927}.

\bibitem[{Engelhardt et~al.(2009)Engelhardt, Martin, Prozorov, Luban, Timco,
  and Winpenny}]{EMP:PRB09}
\bibinfo{author}{L.~Engelhardt}, \bibinfo{author}{C.~Martin},
  \bibinfo{author}{R.~Prozorov}, \bibinfo{author}{M.~Luban},
  \bibinfo{author}{G.~A. Timco}, \bibinfo{author}{R.~E.~P. Winpenny},
  \bibinfo{title}{High-field magnetic properties of the magnetic molecule
  {Cr[sub 10]Cu[sub 2]}}, \bibinfo{journal}{Phys. Rev. B} \bibinfo{volume}{79}
  \bibinfo{eid}{014404}, \doi{\bibinfo{doi}{10.1103/PhysRevB.79.014404}}.

\bibitem[{Todea et~al.(2009)Todea, Merca, B{\"o}gge, Glaser, Engelhardt,
  Prozorov, Luban, and M{\"u}ller}]{TMB:CC09}
\bibinfo{author}{A.~M. Todea}, \bibinfo{author}{A.~Merca},
  \bibinfo{author}{H.~B{\"o}gge}, \bibinfo{author}{T.~Glaser},
  \bibinfo{author}{L.~Engelhardt}, \bibinfo{author}{R.~Prozorov},
  \bibinfo{author}{M.~Luban}, \bibinfo{author}{A.~M{\"u}ller},
  \bibinfo{title}{Polyoxotungstates now also with pentagonal units:
  supramolecular chemistry and tuning of magnetic exchange in {(M)M5}12V30
  Keplerates (M = Mo{,} W)}, \bibinfo{journal}{Chem. Commun.}
  (\bibinfo{year}{2009})
  \bibinfo{pages}{3351--3353}\doi{\bibinfo{doi}{10.1039/B907188A}}.

\bibitem[{Engelhardt and Rainey(2010)}]{EnR:FITMART10}
\bibinfo{author}{L.~Engelhardt}, \bibinfo{author}{C.~Rainey},
  \bibinfo{title}{Tool for Magnetic Analysis Package}, \bibinfo{year}{2010}.

\bibitem[{Borras-Almenar et~al.(2001)Borras-Almenar, Clemente-Juan, Coronado,
  and Tsukerblat}]{BCC:JCC99}
\bibinfo{author}{J.~J. Borras-Almenar}, \bibinfo{author}{J.~M. Clemente-Juan},
  \bibinfo{author}{E.~Coronado}, \bibinfo{author}{B.~S. Tsukerblat},
  \bibinfo{title}{MAGPACK$_1$ A package to calculate the energy levels, bulk
  magnetic properties, and inelastic neutron scattering spectra of high
  nuclearity spin clusters}, \bibinfo{journal}{J. Comp. Chem.}
  \bibinfo{volume}{22} (\bibinfo{year}{2001}) \bibinfo{pages}{985--991},
  \doi{\bibinfo{doi}{10.1002/jcc.1059}}.

\bibitem[{Albuquerque et~al.(2007)Albuquerque, Alet, Corboz, Dayal, Feiguin,
  Fuchs, Gamper, Gull, G{\"u}rtler, Honecker, Igarashi, K{\"o}rner,
  Kozhevnikov, L{\"a}uchli, Manmana, Matsumoto, McCulloch, Michel, Noack,
  Pawlowski, Pollet, Pruschke, Schollw{\"o}ck, Todo, Trebst, Troyer, Werner,
  and Wessel}]{ALPS:JMMM07}
\bibinfo{author}{A.~Albuquerque}, \bibinfo{author}{F.~Alet},
  \bibinfo{author}{P.~Corboz}, \bibinfo{author}{P.~Dayal},
  \bibinfo{author}{A.~Feiguin}, \bibinfo{author}{S.~Fuchs},
  \bibinfo{author}{L.~Gamper}, \bibinfo{author}{E.~Gull},
  \bibinfo{author}{S.~G{\"u}rtler}, \bibinfo{author}{A.~Honecker},
  \bibinfo{author}{R.~Igarashi}, \bibinfo{author}{M.~K{\"o}rner},
  \bibinfo{author}{A.~Kozhevnikov}, \bibinfo{author}{A.~L{\"a}uchli},
  \bibinfo{author}{S.~Manmana}, \bibinfo{author}{M.~Matsumoto},
  \bibinfo{author}{I.~McCulloch}, \bibinfo{author}{F.~Michel},
  \bibinfo{author}{R.~Noack}, \bibinfo{author}{G.~Pawlowski},
  \bibinfo{author}{L.~Pollet}, \bibinfo{author}{T.~Pruschke},
  \bibinfo{author}{U.~Schollw{\"o}ck}, \bibinfo{author}{S.~Todo},
  \bibinfo{author}{S.~Trebst}, \bibinfo{author}{M.~Troyer},
  \bibinfo{author}{P.~Werner}, \bibinfo{author}{S.~Wessel}, \bibinfo{title}{The
  ALPS project release 1.3: Open-source software for strongly correlated
  systems}, \bibinfo{journal}{J. Magn. Magn. Mater.} \bibinfo{volume}{310}
  (\bibinfo{year}{2007}) \bibinfo{pages}{1187 -- 1193}, \doi{\bibinfo{doi}{DOI:
  10.1016/j.jmmm.2006.10.304}}.

\bibitem[{Pollet et~al.(2004)Pollet, Rombouts, Van~Houcke, and
  Heyde}]{PRV:PRE04}
\bibinfo{author}{L.~Pollet}, \bibinfo{author}{S.~M.~A. Rombouts},
  \bibinfo{author}{K.~Van~Houcke}, \bibinfo{author}{K.~Heyde},
  \bibinfo{title}{Optimal Monte Carlo updating}, \bibinfo{journal}{Phys. Rev.
  E} \bibinfo{volume}{70} (\bibinfo{year}{2004}) \bibinfo{pages}{056705},
  \doi{\bibinfo{doi}{10.1103/PhysRevE.70.056705}}.

\bibitem[{Alet et~al.(2005)Alet, Wessel, and Troyer}]{AWT:PRE05}
\bibinfo{author}{F.~Alet}, \bibinfo{author}{S.~Wessel},
  \bibinfo{author}{M.~Troyer}, \bibinfo{title}{Generalized directed loop method
  for quantum Monte Carlo simulations}, \bibinfo{journal}{Phys. Rev. E}
  \bibinfo{volume}{71} (\bibinfo{year}{2005}) \bibinfo{pages}{036706},
  \doi{\bibinfo{doi}{10.1103/PhysRevE.71.036706}}.

\bibitem[{Zheng et~al.(2013)Zheng, Zhang, Long, Huang, M{\"u}ller, Schnack,
  Zheng, and Zheng}]{ZZL:CC13}
\bibinfo{author}{Y.~Zheng}, \bibinfo{author}{Q.-C. Zhang},
  \bibinfo{author}{L.-S. Long}, \bibinfo{author}{R.-B. Huang},
  \bibinfo{author}{A.~M{\"u}ller}, \bibinfo{author}{J.~Schnack},
  \bibinfo{author}{L.-S. Zheng}, \bibinfo{author}{Z.~Zheng},
  \bibinfo{title}{Molybdate templated assembly of Ln$_{12}$Mo$_{4}$-type
  clusters (Ln = Sm{,} Eu{,} Gd) containing a truncated tetrahedron core},
  \bibinfo{journal}{Chem. Commun.} \bibinfo{volume}{49} (\bibinfo{year}{2013})
  \bibinfo{pages}{36--38}, \doi{\bibinfo{doi}{10.1039/C2CC36530H}}.

\bibitem[{Schnack(2010)}]{Sch:DT10}
\bibinfo{author}{J.~Schnack}, \bibinfo{title}{Effects of frustration on
  magnetic molecules: a survey from Olivier Kahn until today},
  \bibinfo{journal}{Dalton Trans.} \bibinfo{volume}{39} (\bibinfo{year}{2010})
  \bibinfo{pages}{4677 -- 4686}, \doi{\bibinfo{doi}{10.1039/B925358K}}.

\bibitem[{Ivanov et~al.(2010)Ivanov, Schnack, Schnalle, Richter, K\"ogerler,
  Newton, Cronin, Oshima, and Nojiri}]{ISS:PRL10}
\bibinfo{author}{N.~B. Ivanov}, \bibinfo{author}{J.~Schnack},
  \bibinfo{author}{R.~Schnalle}, \bibinfo{author}{J.~Richter},
  \bibinfo{author}{P.~K\"ogerler}, \bibinfo{author}{G.~N. Newton},
  \bibinfo{author}{L.~Cronin}, \bibinfo{author}{Y.~Oshima},
  \bibinfo{author}{H.~Nojiri}, \bibinfo{title}{Heat Capacity Reveals the
  Physics of a Frustrated Spin Tube}, \bibinfo{journal}{Phys. Rev. Lett.}
  \bibinfo{volume}{105} (\bibinfo{year}{2010}) \bibinfo{pages}{037206},
  \doi{\bibinfo{doi}{10.1103/PhysRevLett.105.037206}}.

\bibitem[{Trotter(1959)}]{Tro:PAMS59}
\bibinfo{author}{H.~F. Trotter}, \bibinfo{title}{On the product of semi-groups
  of operators}, \bibinfo{journal}{Proc. Amer. Math. Soc.} \bibinfo{volume}{10}
  (\bibinfo{year}{1959}) \bibinfo{pages}{545--551},
  \doi{\bibinfo{doi}{10.1090/S0002-9939-1959-0108732-6}}.

\bibitem[{Suzuki(1976)}]{Suz:CMP76}
\bibinfo{author}{M.~Suzuki}, \bibinfo{title}{Generalized Trotter's formula and
  systematic approximants of exponential operators and inner derivations with
  applications to many-body problems}, \bibinfo{journal}{Commun. Math. Phys.}
  \bibinfo{volume}{51} (\bibinfo{year}{1976}) \bibinfo{pages}{183--190},
  \doi{\bibinfo{doi}{10.1007/BF01609348}}.

\bibitem[{Suzuki(1977)}]{Suz:CMP77}
\bibinfo{author}{M.~Suzuki}, \bibinfo{title}{On the convergence of exponential
  operators -- the Zassenhaus formula, BCH formula and systematic
  approximants}, \bibinfo{journal}{Commun. Math. Phys.} \bibinfo{volume}{57}
  (\bibinfo{year}{1977}) \bibinfo{pages}{193--200},
  \doi{\bibinfo{doi}{10.1007/BF01614161}}.

\bibitem[{Taft et~al.(1994)Taft, Delfs, Papaefthymiou, Foner, Gatteschi, and
  Lippard}]{TDP:JACS94}
\bibinfo{author}{K.~L. Taft}, \bibinfo{author}{C.~D. Delfs},
  \bibinfo{author}{G.~C. Papaefthymiou}, \bibinfo{author}{S.~Foner},
  \bibinfo{author}{D.~Gatteschi}, \bibinfo{author}{S.~J. Lippard},
  \bibinfo{title}{[Fe(OMe)$_2$(O$_2$CCH$_2$Cl)]$_{10}$, a molecular ferric
  wheel}, \bibinfo{journal}{J.~Am. Chem. Soc.} \bibinfo{volume}{116}
  (\bibinfo{year}{1994}) \bibinfo{pages}{823}.

\bibitem[{King et~al.(2006)King, Stamatatos, Abboud, and Christou}]{KSA:ACIE06}
\bibinfo{author}{P.~King}, \bibinfo{author}{T.~C. Stamatatos},
  \bibinfo{author}{K.~A. Abboud}, \bibinfo{author}{G.~Christou},
  \bibinfo{title}{Reversible Size Modification of Iron and Gallium Molecular
  Wheels: A Ga$_{10}$ Gallic Wheel and Large Ga$_{18}$ and Fe$_{18}$ Wheels},
  \bibinfo{journal}{Angew. Chem. Int. Ed.}
  \bibinfo{volume}{45}~(\bibinfo{number}{44}) (\bibinfo{year}{2006})
  \bibinfo{pages}{7379--7383}.

\bibitem[{Schnack and Luban(2000)}]{ScL:PRB00}
\bibinfo{author}{J.~Schnack}, \bibinfo{author}{M.~Luban},
  \bibinfo{title}{Rotational modes in molecular magnets with antiferromagnetic
  {H}eisenberg exchange}, \bibinfo{journal}{Phys. Rev. B} \bibinfo{volume}{63}
  (\bibinfo{year}{2000}) \bibinfo{pages}{014418},
  \doi{\bibinfo{doi}{10.1103/PhysRevB.63.014418}}.

\bibitem[{Honecker et~al.(2004)Honecker, Schulenburg, and Richter}]{HSR:JP04}
\bibinfo{author}{A.~Honecker}, \bibinfo{author}{J.~Schulenburg},
  \bibinfo{author}{J.~Richter}, \bibinfo{title}{Magnetization plateaus in
  frustrated antiferromagnetic quantum spin models}, \bibinfo{journal}{J.
  Phys.: Condens. Matter} \bibinfo{volume}{16} (\bibinfo{year}{2004})
  \bibinfo{pages}{S749}.

\bibitem[{Micotti et~al.(2006)Micotti, Furukawa, Kumagai, Carretta,
  Lascialfari, Borsa, Timco, and Winpenny}]{MFK:PRL06}
\bibinfo{author}{E.~Micotti}, \bibinfo{author}{Y.~Furukawa},
  \bibinfo{author}{K.~Kumagai}, \bibinfo{author}{S.~Carretta},
  \bibinfo{author}{A.~Lascialfari}, \bibinfo{author}{F.~Borsa},
  \bibinfo{author}{G.~A. Timco}, \bibinfo{author}{R.~E.~P. Winpenny},
  \bibinfo{title}{Local spin moment distribution in antiferromagnetic molecular
  rings probed by NMR}, \bibinfo{journal}{Phys. Rev. Lett.}
  \bibinfo{volume}{97} (\bibinfo{year}{2006}) \bibinfo{pages}{267204},
  \urlprefix\url{http://dx.doi.org/10.1103/PhysRevLett.97.267204}.

\bibitem[{Furukawa et~al.(2007)Furukawa, Nishisaka, Kumagai, K\"ogerler, and
  Borsa}]{FNK:PRB07}
\bibinfo{author}{Y.~Furukawa}, \bibinfo{author}{Y.~Nishisaka},
  \bibinfo{author}{K.-i. Kumagai}, \bibinfo{author}{P.~K\"ogerler},
  \bibinfo{author}{F.~Borsa}, \bibinfo{title}{Local spin moment configuration
  in the frustrated $s=1/2$ Heisenberg triangular antiferromagnet V15
  determined by NMR}, \bibinfo{journal}{Phys. Rev. B} \bibinfo{volume}{75}
  (\bibinfo{year}{2007}) \bibinfo{pages}{220402},
  \doi{\bibinfo{doi}{10.1103/PhysRevB.75.220402}}.

\bibitem[{Prinz et~al.(2010)Prinz, Kuepper, Taubitz, Raekers, Biswas,
  Weyherm{\"u}ller, Uhlarz, Wosnitza, Schnack, Postnikov, Schr{\"o}der, George,
  Neumann, , and Chaudhuri}]{PKT:IC10}
\bibinfo{author}{M.~Prinz}, \bibinfo{author}{K.~Kuepper},
  \bibinfo{author}{C.~Taubitz}, \bibinfo{author}{M.~Raekers},
  \bibinfo{author}{B.~Biswas}, \bibinfo{author}{T.~Weyherm{\"u}ller},
  \bibinfo{author}{M.~Uhlarz}, \bibinfo{author}{J.~Wosnitza},
  \bibinfo{author}{J.~Schnack}, \bibinfo{author}{A.~V. Postnikov},
  \bibinfo{author}{C.~Schr{\"o}der}, \bibinfo{author}{S.~J. George},
  \bibinfo{author}{M.~Neumann}, , \bibinfo{author}{P.~Chaudhuri},
  \bibinfo{title}{A Star-Shaped Heteronuclear
  Cr$^{\text{III}}$Mn$^{\text{II}}_3$ Species and Its Precise Electronic and
  Magnetic Structure: Spin Frustration Studied by X-Ray Spectroscopic,
  Magnetic, and Theoretical Methods}, \bibinfo{journal}{Inorg. Chem.}
  \bibinfo{volume}{49} (\bibinfo{year}{2010}) \bibinfo{pages}{2093--2102},
  \doi{\bibinfo{doi}{10.1021/ic9012119}}.

\end{thebibliography}

\end{document}